\documentclass[10pt,twocolumn,aps,prb]{revtex4-1}
\usepackage{graphicx}
\usepackage{amsmath}

\begin{document}
\title{Spin injection from a half-metal at finite temperatures}
\author{K.\ D.\ Belashchenko}
\author{J.\ K.\ Glasbrenner}
\author{A.\ L.\ Wysocki}
\affiliation{Department of Physics and Astronomy and Nebraska Center for Materials and Nanoscience, University of Nebraska, Lincoln, Nebraska 68588, USA}

\date{\today}

\begin{abstract}
Spin injection from a half-metallic electrode in the presence of thermal spin disorder is analyzed using a combination of random matrix theory, spin-diffusion theory, and explicit simulations for the tight-binding $s$-$d$ model. It is shown that efficient spin injection from a half-metal is possible as long as the effective resistance of the normal metal does not exceed a characteristic value, which does not depend on the resistance of the half-metallic electrode, but is rather controlled by spin-flip scattering at the interface. This condition can be formulated as $\alpha\lesssim l/l^N_{sf}T_c^{-1}$, where $\alpha$ is the relative deviation of the magnetization from saturation, $l$ and $l^N_{sf}$ the mean-free path and the spin-diffusion length in the non-magnetic channel, and $T_c$ the transparency of the tunnel barrier at the interface (if present). The general conclusions are confirmed by tight-binding $s$-$d$ model calculations. A rough estimate suggests that efficient spin injection from true half-metallic ferromagnets into silicon or copper may be possible at room temperature across a transparent interface.
\end{abstract}

\maketitle

\section{Introduction}

Many spintronic devices depend on the injection, manipulation, and detection of spin-polarized currents in semiconductors or normal metals.\cite{Zutic,Johnson-chapter,Jansen-r} Spin injection can also be utilized as a tool to probe the spectroscopic properties of strongly correlated and spin-orbit-coupled systems.\cite{Zutic,Cheuk} Thus, understanding the mechanisms of spin injection is of interest for a variety of fundamental and practical applications. Basic theory of spin injection across an F/N (ferromagnet/normal metal) interface in the linear response regime was worked out by Johnson and Silsbee.\cite{Johnson} The spin polarization of the injected current may be conveniently expressed as \cite{Rashba,Zutic}
\begin{equation}\label{Pj}
P_j=\frac{P_\sigma r_F + P_\Sigma r_c}{r_F+r_N+r_c},
\end{equation}
where $P_\sigma=(\sigma_\uparrow-\sigma_\downarrow)/(\sigma_\uparrow+\sigma_\downarrow)$, $\sigma_\uparrow$ and $\sigma_\downarrow$ are the spin-resolved conductivities of the ferromagnetic electrode, $P_\Sigma$ is defined similar to $P_\sigma$ for the spin-dependent interface conductance, while $r_F$ and $r_N$ are effective resistances of the ferromagnet and normal metal, respectively. The effective interface resistance is denoted $r_c$. For the
ferromagnet $r_F=(\sigma_\uparrow+\sigma_\downarrow)l^F_{sf}/(4\sigma_\uparrow\sigma_\downarrow)$ and for the normal metal $r_N=l^N_{sf}/\sigma_N$, where $l^F_{sf}$ and $l^N_{sf}$ are the spin-diffusion lengths in the ferromagnet and in the normal metal. The quantities $r_F$ and $r_N$ are called effective resistances. The expression (\ref{Pj}) is valid under the assumptions of the two-current model, \cite{Mott} i.\ e.\ when the spin-diffusion lengths are much longer than the mean-free paths.\cite{VF} Note that nonlinear effects in bipolar semiconducting junctions \cite{bipolar,Zutic02} can not be described within the linear-response theory and are beyond the scope of the present consideration.

Spin injection from a ferromagnet into a semiconductor is subject to the so-called conductivity mismatch problem.\cite{Schmid} For a typical choice of materials we have $l^N_{sf}\gg l^F_{sf}$ and $\sigma_N\ll\sigma_\uparrow\sigma_\downarrow/(\sigma_\uparrow+\sigma_\downarrow)$. This implies that $r_F\ll r_N$, and if the interface resistance $r_c$ is also small compared to $r_N$, the injected current is unpolarized, $P_j\ll 1$. In order to circumvent this problem, one can introduce a highly resistive, spin-selective barrier at the interface, such as a naturally occurring Schottky barrier or an artificially inserted tunnel junction.\cite{Rashba} According to Eq.\ (\ref{Pj}), large $r_c$ (comparable to or greater than $r_N$) combined with appreciable $P_\Sigma$ results in a finite $P_j$. Efficient spin injection into GaAs and Si from transition-metal electrodes was successfully achieved based on this principle. \cite{Hanbicki,Crowell,Jonker,Jansen}

The situation can be visualized with the help of an effective resistor circuit, such that in each spin channel $s$ the ferromagnet and the normal metal have resistances $l^F_{sf}/\sigma_s$ and $2l^N_{sf}/\sigma_N$, respectively, and the two spin channels are connected in parallel.\cite{JEPP} This effective circuit correctly reproduces both the spin polarization of the current near the interface and the resistance of the junction in excess of what would be measured if the interface were replaced by a node in the circuit.\cite{JEPP}

Half-metallic ferromagnets \cite{Groot} are conducting in one spin channel and insulating in the other, which makes them attractive candidates as electrode materials for spintronic devices.\cite{Zutic}
The situation at zero temperature is simple, as there is only one conducting channel (``spin up''), and the injected current should be fully spin-polarized.
Many materials, particularly among Heusler compounds, have been theoretically predicted using band structure calculations to be half-metallic,\cite{Katsnelson} although reliable experimental confirmation is often complicated by surface effects.\cite{Dowben} High magnetoresistance values were achieved in magnetic tunnel junctions \cite{Sakuraba,Marukame,Ishikawa,Sukegawa} and spin valves \cite{Nakatani} with epitaxial Co-based Heusler-alloy electrodes. Large nonlocal spin signals, 10 times higher compared to conventional electrodes, were also demonstrated in lateral spin valves with transparent Ohmic interfaces.\cite{Bridoux,Kimura,Hamaya} In all of these experiments the spin signal is considerably reduced at room temperature but remains appreciable. Further, Ramsteiner \emph{et al.} \cite{Ramsteiner} demonstrated spin injection from Co$_2$FeSi into an (Al,Ga)As light-emitting diode (LED) structure with an efficiency of at least 50\%. Based on their device design, they concluded that the Schottky barrier can not be present at the interface and argued that the observed efficient spin injection ``casts doubt onto the common belief that tunneling is a prerequisite for spin injection from a metal into a semiconductor.''

So far the theoretical analysis of spintronic devices with half-metallic electrodes was based \cite{Bridoux,Kimura,Hamaya} on the standard spin-diffusion model, \cite{Johnson,VF,Rashba,Takahashi,Zutic} which assumes the existence of two weakly coupled conducting channels in each material.\cite{Mott,VF} However, these validity criteria are not satisfied in true half-metals, and our present goal is to develop an appropriate formalism for such devices, which should include the effects of thermal spin fluctuations at finite temperatures.
Due to these fluctuations, the electron wavefunctions lose their pure spin character, and the density of states (DOS) in a half-metal acquires non-zero projection onto the ``spin down'' channel, which is gapped at zero temperature. However, this state can be viewed as a small perturbation of the fully collinear spin state by fluctuating transverse magnetic fields, so that the local spin direction for all electronic eigenstates is fluctuating within a narrow cone around the magnetization direction. In other words, the number of eigenstates is not doubled, but rather they acquire a small spin-down component. In this situation one can not apply the two-current model, in which independent distribution functions are introduced for spin-up and spin-down electrons, and the concept of the spin-diffusion length also becomes meaningless. Therefore, Eq.\ (\ref{Pj}) can not be directly applied to spin injection from a half-metallic electrode at $T\neq0$.

In the following, we analyze the spin injection from a half-metallic electrode in the linear response regime but without making the assumptions of the two-current model leading to Eq.\ (\ref{Pj}). We start with general considerations in Section \ref{general} and then proceed to analyze the elastic scattering region using the random matrix theory in Section \ref{scattering}. Here we derive the formula for spin injection efficiency, which is similar to (\ref{Pj}) but with the effective resistance $\tilde r_F$ being controlled by spin-flip scattering probabilities at the interface. A generalized statement of the conductivity mismatch follows from the unitarity of the scattering matrix. Based on these results, we then describe the half-metallic spin-injection system within the spin-diffusion theory in Section \ref{DiffTh}. The formula for spin-injection efficiency is  generalized in a natural way to the case of finite spin-diffusion length in the normal region. In Section \ref{properties} we discuss the behavior of spin-injection efficiency, and finally in Section \ref{TB} we support our conclusions with explicit tight binding $s$-$d$ model calculations. The conclusions are summarized in Section \ref{concl}.

\section{Half-metal at a finite temperature}\label{general}

The electronic structure of a half-metal at $T=0$ has a band gap in one of the spin channels. If we now consider a thermal fluctuation resulting in a small canting of individual local spin moments, we can imagine, on the level of an the self-consistent field theory in the localized basis, that the effective fields on different atomic sites have been rigidly rotated off of the magnetization axis by small angles. (The tight-binding representation is assumed for simplicity and is not essential for the physical argument.) This is a common approach to spin fluctuations within the noncollinear density functional theory, whereby the spin moments are assumed to fluctuate adiabatically slowly compared with electron hopping times.\cite{Gyorffy} This approximation is justified by the fact that typical times associated with magnon dynamics are much longer compared to the electron momentum relaxation time. The Hamiltonian of the system with such adiabatic spin fluctuation can then be represented as \cite{Skomski}
\begin{equation}\label{HSDR}
    H\{\hat n_i\}=\sum_i U(\hat n_i) H_i U^+(\hat n_i)+ K
\end{equation}
where $\hat n_i$ is the unit vector parallel to the spin moment on site $i$, $H_i$ is the on-site contribution to the Hamiltonian from site $i$, and $K$ is the spin-diagonal kinetic (hopping) part.
We can now make a unitary transformation to the new ``rotated'' local basis in which $H_i$ is diagonal, which is effected by unitary matrices $U(\hat n_i)$. In this new basis the Hamiltonian is
\begin{equation}\label{Hrot}
    \tilde H\{\hat n_i\}=\sum_i H_i + \sum_{ij} U^+(\hat n_i)K_{ij}U(\hat n_j) .
\end{equation}
At zero temperature there are only states of a particular spin (say, ``up'') near the Fermi level. Deviation of the unitary matrices in (\ref{Hrot}) from unity at $T\neq0$ introduces hybridization between local spin-up and spin-down states, as well as some randomness in the spin-conserving hopping matrix elements. The Bloch states near the Fermi level acquire a small admixture of spin-down character, and the DOS in the global basis acquires a spin-down component, but no new Bloch states appear near the Fermi level. This means that the bulk of a half-metal at finite (but low) temperatures can be treated as having one effective spin channel. This situation is qualitatively different from a conventional ferromagnet with two independent spin channels, even if they have very different resistivities. While the conventional ferromagnet has two independent occupation functions and chemical potentials for the two spin channels, a half-metal has only one. Transport across an interface with a normal metal is discussed in the subsequent sections.

\section{Spin injection in the scattering formalism}\label{scattering}

A spin injection device can be analyzed by treating the F/N interface as an elastic scattering region embedded between diffusive regions, and by matching the solution of the scattering problem with the solution of the spin-diffusion equation. For a conventional two-channel ferromagnetic electrode the well-known result is given by Eq.\ (\ref{Pj}). However, as we have argued in Section \ref{general}, the two-current model is inapplicable for a half-metallic electrode. The purpose of this section is to understand the role of spin coherence for spin injection from the half-metal at finite temperatures, i.\ e.\ in the presence of spin disorder.

Since the half-metal, as argued above, has only one effective spin channel, there is no analog of the spin-diffusion length for it. Therefore, inelastic scattering should not affect the properties of spin injection, and we may treat the whole half-metallic electrode as an elastic scatterer. As we will see in the next section, matching with the solution of the spin-diffusion equation in the normal metal should simply replace the resistance of the normal region by its effective resistance $r_N=\rho_N l^N_{sf}$. Therefore, we first consider the entire spin-injection device disregarding inelastic scattering altogether.

An elastic spin-injection device can be considered in the formalism of the scattering theory. We assume that the F-N device is connected on both sides to equilibrium reservoirs via ideal Landauer leads. Apart from these
leads we introduce an auxiliary lead in the N region at such distance from the interface (a few mean-free paths) that the quantum interference effects occurring at the interface are left entirely on the left-hand side of this lead. Each of the two regions can be described by a scattering matrix:
\begin{equation}\label{SF}
    \hat S_F=\left(\begin{array}{cc}
    \hat r_1 & \hat t'_1 \\ \hat t_1 & \hat r'_1 \\
    \end{array}\right) \quad,\quad \hat S_N=\left(\begin{array}{cc}
    \hat r_2 & \hat t'_2 \\ \hat t_2 & \hat r'_2 \\
    \end{array}\right)
\end{equation}
where, in the standard way, the matrix $\hat t_1$ contains amplitudes for transmission from the conducting channels of the left electrode across F into the conducting channels of the fictitious lead, and similarly for the other
subblocks. At this point we allow the F region to have an arbitrary magnetic configuration. The transmission matrix $\hat t$ of the entire F-N junction is
\begin{equation}\label{t}
    \hat t=\hat t_2(1-\hat r'_1\hat r_2)^{-1}\hat t_1 .
\end{equation}

The charge and spin currents flowing across the junction are proportional, respectively, to $C$ and $C_s$:
\begin{equation}\label{j}
   C=\mathop{\mathrm{Tr}}\hat t\hat t^+\quad,\quad C_s=\mathop{\mathrm{Tr}}\hat\sigma_z \hat t\hat t^+
\end{equation}
and we are interested in the spin polarization $P_j=C_s/C$. Since we are considering the junction as an elastic scattering region, the spin current in the normal region is conserved. There is no loss of generality from singling out the $z$ axis, because its direction is unspecified.

Following the approach of Waintal \emph{et al.},\cite{Waintal} we now introduce the polar decomposition\cite{Mello} of the matrix $\hat S_N$:
\begin{equation}\label{polar}
    \hat S_N=\left(\begin{array}{cc}
    \hat U & 0 \\ 0 & \hat V' \\
    \end{array}\right)
    \left(\begin{array}{cc}
    \sqrt{1-T} & i\sqrt{T} \\ i\sqrt{T} & \sqrt{1-T} \\
    \end{array}\right)
    \left(\begin{array}{cc}
    \hat U' & 0 \\ 0 & \hat V \\
    \end{array}\right)
\end{equation}
where $T$ is the matrix of the eigenvalues of $\hat t_2 \hat t_2^+$, while $\hat U$, $\hat U'$, $\hat V$, and $\hat V'$ are unitary matrices, which are
all diagonal in spin space. Since the fictitious node can be introduced at a sufficient distance from the surface to eliminate all quantum interference
effects, we can safely use the isotropic approximation, \cite{Beenakker} i.\ e.\ assume that the spatial factors of the unitary matrices $\hat U$, $\hat
U'$, $\hat V$, and $\hat V'$ are distributed uniformly in the unitary group. We substitute (\ref{t}) in (\ref{j}), use (\ref{polar}) for $\hat t_2$ and
$\hat r'_2$, and integrate over the unitary ensemble. This integration is easily performed using the method of Ref.\ \onlinecite{Brouwer} to the
leading order in the number of conducting channels in the leads. In this leading order, each unitary matrix is matched to its own conjugate, resulting
in a ladder diagram. \cite{Brouwer} Averaging over the eigenvalues of $\hat t_2 \hat t_2^+$ is performed simultaneously. The result can be written in this form:
\begin{equation}\label{js}
    C_s=\sum_{\lambda\mu}\sigma^z_{\lambda}\left[1-\hat R(1-T_N)\right]^{-1}_{\lambda\mu}T_{\mu} T_N .
\end{equation}
Here $\lambda$ and $\mu$ denote a pair of spin indices, $\sigma^z_\lambda$ is the $\hat \sigma_z$ matrix written as a vector $(1,0,0,-1)$, $T_N$ is the
probability of transmission through the N region, $\hat R$ is a $4\times4$ matrix\cite{Waintal} with elements
\begin{equation}\label{Rmat}
    R_{\lambda\mu}\equiv
    R_{\sigma\sigma',ss'}=\frac{1}{N_{ch}}\sum_{mn}(r'_1)_{m\sigma,ns}(r'_1)^*_{m\sigma',ns'}
\end{equation}
where $m$, $n$ enumerate the $N_{ch}$ conducting channels in the auxiliary lead, and $T_\mu$ is a 4-vector with elements
\begin{equation}\label{tvec}
    T_{\mu}\equiv T_{ss'}=\frac{1}{N_{ch}}\sum_{mn\sigma}(t_1)_{ms,n\sigma}(t_1)^*_{ms',n\sigma}.
\end{equation}
In this last expression $n\sigma$ labels the channels of the lead feeding the F region.
The expression for $C$ is obtained from (\ref{js}) replacing $\sigma^z_\lambda$ by a 4-vector representation of the unit matrix
$1_\lambda=(1,0,0,1)$. Note that $T_\mu=\sum_\lambda T_{\mu\lambda} 1_\lambda$, where $T_{\mu\lambda}$ is defined as $R_{\mu\lambda}$ but with matrix elements of $t_1$ instead of $r^\prime_1$.\cite{note-hat}

The unitarity of $\hat S_F$ requires that $\hat t_1\hat t_1^+ + \hat r'\hat r'^+=1$. This condition implies that $T_\lambda=\sum_\mu(\delta_{\lambda\mu}- R_{\lambda\mu})1_\mu$. Substituting this in (\ref{js}) we find that $C_s$ vanishes to first order in $T_N$. Since the charge current is proportional to $T_N$, this leads to $P_j\to0$ at $T_N\to0$. In the limit of a two-channel device with weak coupling between the channels, this result reduces to the conductivity mismatch obstacle for spin injection. \cite{Schmid} However, our result is more general, because it is valid for any magnetic structure of the F region and for any choice of the $z$ axis. The only exception is the case of a half-metal at $T=0$ with no spin-flip scattering at the interface, for which the only non-zero component of $T_\mu$ is $T_{\uparrow\uparrow}$ (in the reference frame where the $z$ axis is aligned with the magnetization). In this exceptional case we obviously have $P_j=1$ at any $T_N$.

Let us now find the spin polarization $P_j$ for a finite $T_N$, assuming that the electrode is an axially symmetric (i.\ e.\ collinear) magnet with spin disorder. For a macroscopic interface, the summation over the conducting channels automatically averages $R_{\lambda\mu}$ and $T_\mu$ in (\ref{js}) over the spin disorder ensemble. This self-averaging does not necessarily occur in a point-contact, in which case an additional averaging over
spin disorder configuration is required for the spin current (\ref{js}) and its charge counterpart. Re-expanding the inverse matrix in (\ref{js}), we can obtain a series of terms describing multiple scatterings at the interface. Since the electron scatters repeatedly from the same spin disorder configuration, the averages of the matrix products do not decouple. However, since correlations between successive scattering events do not change the asymptotic behavior of $P_j$, it is a reasonable approximation to replace $\hat R$ and $T_\mu$ by their averages $\langle\hat R\rangle$ and $\langle T_\mu\rangle$ even for a point contact.

Let us assume that spin-orbit coupling at the surface is weak, and that all spin-flip processes are dominated by spin-disorder scattering. Then
the matrices $\langle\hat R\rangle$ and $\langle T_\mu\rangle$ should be invariant with respect to rotation in spin space around the magnetization axis.
This condition implies that $\langle T_{\uparrow\downarrow}\rangle$ and $\langle T_{\downarrow\uparrow}\rangle$ vanish, along with all elements
$\langle R_{\sigma\sigma',ss'}\rangle$ with $\sigma-\sigma'\ne s-s'$. The $\langle\hat R\rangle$ matrix is thus block-diagonal. From the structure of
(\ref{js}) it is clear that we are only interested in the $2\times2$ block spanned by indices 1 and 4. (The 22 and 33 diagonal elements represent the
spin-mixing conductance,\cite{Brataas} which turns out to be irrelevant to the problem at hand.) As seen from (\ref{Rmat}), the diagonal elements of
this block are the total spin-conserving reflection probabilities for spin-up and spin-down electrons $R_\uparrow=\langle
R_{\uparrow\uparrow,\uparrow\uparrow}\rangle$ and $R_\downarrow=\langle R_{\downarrow\downarrow,\downarrow\downarrow}\rangle$, while the off-diagonal
elements are the total spin-flip reflection probabilities $R_{\uparrow\downarrow}=\langle R_{\uparrow\uparrow,\downarrow\downarrow}\rangle$. Reciprocity requires that
$\langle R_{\uparrow\uparrow,\downarrow\downarrow}\rangle=\langle R_{\downarrow\downarrow,\uparrow\uparrow}\rangle$.
Let us also denote $T_\uparrow=T_{\uparrow\uparrow}$ and $T_\downarrow=T_{\downarrow\downarrow}$. (Note that $T_\uparrow$ and $T_\downarrow$ include both spin-conserving and spin-flip processes.)
We can now calculate the spin polarization from (\ref{js}):
\begin{equation}\label{pjfromt}
    P_j=\frac{P_t \tilde r_F}{\tilde r_F+r_N}
\end{equation}
where $P_t=(T_\uparrow-T_\downarrow)/(T_\uparrow+T_\downarrow)$,
$r_NG_N=(1-T_N)/(2T_N)$,
\begin{equation}\label{rf}
\frac{1}{4G_N\tilde r_F}=\frac{T_\uparrow
T_\downarrow}{T_\uparrow+T_\downarrow}+R_{\uparrow\downarrow} ,
\end{equation}
and $G_N=(e^2/h)N_{ch}$. Note that $P_t$ is the spin polarization of the current injected in the auxiliary lead if the N region is detached from it.

The expression (\ref{pjfromt}) includes the effects of spin disorder, but not the effects of inelastic spin relaxation. For a conventional (not half-metallic) electrode, spin relaxation must be included on both sides of the junction. In the presence of spin-flip processes at the interface, the solution of the spin-diffusion equations becomes rather complicated \cite{Rashba2} even if spin-flip reflection $R_{\uparrow\downarrow}$ is neglected.
The situation is simpler in the case of a half-metallic electrode, because inelastic spin relaxation should only be included in the N region. In the next section we will see that in this case the elastic resistance of the normal region $r_N$ in (\ref{pjfromt}) should simply be replaced by its effective resistance.

Note that interfacial spin-flip scattering due to spin-orbit interaction was studied in some detail for metallic N/N and F/N interfaces. \cite{Bass,Nguyen} Temperature-dependent interfacial spin-flip scattering in the presence of non-equilibrium spin accumulation was suggested as a source of asymmetric response in a nonlocal spin valve. \cite{Garzon}

\section{Semiclassical theory}\label{DiffTh}

In the previous section we found that under rather general assumptions the scattering at the interface between the half-metal and the normal metal is described completely by spin-dependent transmission probabilities and the spin-flip reflection probability on the normal metal side. The effects of spin coherence are effectively eliminated by spin disorder averaging. We can therefore use the standard semi-classical treatment, taking into account that the half-metal has only one spin channel, and incorporating spin-flip scattering at the interface. Apart from giving a complementary picture of spin injection, this treatment confirms the expectation about the role of the spin-diffusion length in the normal metal and shows the invariance of the results with respect to the location of the left lead.

Instead of treating the whole F/N device as an elastic scatterer, we now consider only the interfacial F/N region (a few mean-free paths on both sides) embedded between infinite diffusive regions. The half-metallic (F) region carries only one spin channel (even at finite temperature), but the N region has two channels. Similarly to Rashba's treatment of the F-N junction with spin-flip transmission at the interface,\cite{Rashba2} the interface is assigned the spin-flip conductance $\Sigma_{\uparrow\downarrow}$ in addition to the spin-conserving $\Sigma_{\uparrow\uparrow}$. In addition to these terms, we also need to introduce spin relaxation in the normal metal due to spin-flip scattering at the interface. Physically, even if the F electrode is insulating, the spin accumulation in the normal metal can relax through interfacial spin-flip scattering. Introducing the appropriate electrochemical potential drops at the interface, the spin-dependent currents on the normal metal side of the interface can be written as follows:
\begin{align}\label{diffusion}
    j^N_\uparrow(0)=\Sigma_{\uparrow\uparrow}(\zeta^N_\uparrow-\zeta^F)+\tilde R_{\uparrow\downarrow}(\zeta^N_\uparrow-\zeta^N_\downarrow)\\
    j^N_\downarrow(0)=\Sigma_{\uparrow\downarrow}(\zeta^N_\downarrow-\zeta^F)+\tilde R_{\uparrow\downarrow}(\zeta^N_\downarrow-\zeta^N_\uparrow)
    \label{diffusion2}
\end{align}
where the new term is the one with $\tilde R_{\uparrow\downarrow}$. Matching with the solution of the spin-diffusion equation can be worked out in the usual way.\cite{Rashba2} The terms containing bulk
conductivity in the F region drop out, and after some algebra we reproduce Eq.\ (\ref{pjfromt}) with $r_N$ now being the effective resistance $\rho_N l^N_{sf}$ (as anticipated), $P_t$ replaced by $P_\Sigma=(\Sigma_{\uparrow\uparrow}-\Sigma_{\uparrow\downarrow})/(\Sigma_{\uparrow\uparrow}+\Sigma_{\uparrow\downarrow})$, and
\begin{equation}\label{rfa}
\frac{1}{4\tilde
r_F}=\frac{\Sigma_{\uparrow\uparrow}\Sigma_{\uparrow\downarrow}}{\Sigma_{\uparrow\uparrow}+\Sigma_{\uparrow\downarrow}}+\tilde R_{\uparrow\downarrow}.
\end{equation}

Eq.\ (\ref{rfa}) is equivalent to (\ref{rf}) with the replacement $G_NT_\uparrow\to\Sigma_{\uparrow\uparrow}$, $G_NT_\downarrow\to\Sigma_{\uparrow\downarrow}$, and $G_NR_{\uparrow\downarrow}\to\tilde R_{\uparrow\downarrow}$. At first sight, there is a discrepancy, because $\Sigma_{\uparrow\uparrow}$ and $\Sigma_{\uparrow\downarrow}$ are the interface conductances while $T_s$ are the total transmission probabilities of the entire half-metallic electrode. However, these expressions are, in fact, consistent, because $\tilde r_F$ is invariant with respect to the choice of the boundary of the interface region at which the chemical potential $\zeta_F$ is evaluated. In order to see this, let us rewrite Eq.\ (\ref{diffusion})-(\ref{diffusion2}) for the same F-N junction with a different choice of this boundary and denote the new chemical potential (at that boundary) by $\zeta^F_0$. This can be viewed as a simple redefinition of the thickness of the interface region. The chemical potentials on the normal side of the interface, however, are evaluated at the same point. The conductance and
reflectance parameters corresponding to the new choice of $\zeta^F_0$ will be denoted $\Sigma^0_{\uparrow\uparrow}$, $\Sigma^0_{\uparrow\downarrow}$, and $\tilde R^0_{\uparrow\downarrow}$.

For a half-metallic electrode the spin polarization of the current injected into the N region under the condition $\zeta^N_\uparrow=\zeta^N_\downarrow$ is determined by the ratio $\alpha=\Sigma_{\uparrow\downarrow}/\Sigma_{\uparrow\uparrow}$, which should depend only on temperature. Therefore, $\Sigma_{\uparrow\downarrow}/\Sigma_{\uparrow\uparrow}=\Sigma^0_{\uparrow\downarrow}/\Sigma^0_{\uparrow\uparrow}$. The charge current is
\begin{equation}\label{redef}
    j=\Sigma_{\uparrow\uparrow}(\xi_\uparrow+\alpha\xi_\downarrow)=\Sigma_F\Delta\zeta_F
\end{equation}
where we denoted $\xi_s=\zeta^N_s-\zeta^F$ and $\Delta\zeta_F=\zeta^F-\zeta^F_0$, and $\Sigma_F$ is the conductance of the
half-metallic region between the points where $\zeta^F$ and $\zeta^F_0$ are evaluated.

Equating the two different expressions for the same spin-dependent currents
(\ref{diffusion})-(\ref{diffusion2}), we can write
\begin{align}\label{eqn1}
    &(\Sigma^0-\Sigma)\xi_\uparrow + \Sigma^0\Delta\zeta_F+(\tilde R^0-\tilde R)(\xi_\uparrow-\xi_\downarrow)=0\\
    \alpha&(\Sigma^0-\Sigma)\xi_\downarrow + \alpha\Sigma^0\Delta\zeta_F-(\tilde R^0-\tilde R)(\xi_\uparrow-\xi_\downarrow)=0
    \label{eqn2}
\end{align}
where we simplified the notation by dropping indices: $\Sigma=\Sigma_{\uparrow\uparrow}$, $\tilde R=\tilde R_{\uparrow\downarrow}$,
and similarly for $\Sigma^0$ and $\tilde R^0$. Furthermore, substituting
$\Delta\zeta_F$ from (\ref{redef}), we obtain a system of two linear
homogeneous equations for $\xi_\uparrow$ and $\xi_\downarrow$. For this
system to have a solution, the determinant of the coefficient matrix should
vanish, which leads to
\begin{align}\label{det}
   \frac{\alpha}{1+\alpha}(\Sigma^0-\Sigma)+\tilde R_0-\tilde R=0 .
\end{align}
This expression implies that $\tilde r_F$ defined in (\ref{rfa}) does not depend on the definition of the boundary of the interface region on the half-metallic side. Physically, this property follows from the unitarity of the scattering matrix and can not be satisfied without introducing the spin-flip reflection terms in (\ref{diffusion})-(\ref{diffusion2}).

\begin{figure}
\begin{center}
\includegraphics[width=0.35\textwidth]{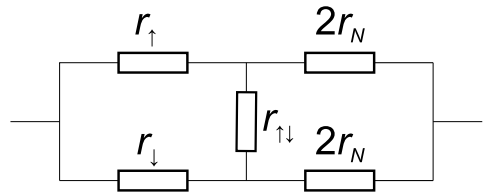}
\end{center}
\caption{Equivalent resistor circuit for spin injection from a half-metal.} \label{resistors}
\end{figure}

Spin injection from a half-metallic electrode may be schematically represented by the equivalent resistor circuit shown in Fig.\ \ref{resistors}. The resistances are defined as $r_{\uparrow\downarrow}=1/\tilde R_{\uparrow\downarrow}$, $r_\uparrow = 1/\Sigma_{\uparrow\uparrow}$, and $r_\downarrow = 1/\Sigma_{\uparrow\downarrow}$. Interestingly, the effective resistance $4\tilde r_F$ in (\ref{rfa}) can measured between the terminals of $r_{\uparrow\downarrow}$ if the N part of the circuit is disconnecting and the left terminal is left open. $P_\Sigma$ is given by $(r_\downarrow-r_\uparrow)/(r_\downarrow+r_\uparrow)$. The physical location of the left terminal of the circuit can be selected anywhere inside the half-metal. According to the arguments presented above, the change of this location redefines the three resistances $r_\uparrow$, $r_\downarrow$, $r_{\uparrow\downarrow}$ while leaving $r_\uparrow/r_\downarrow\sim\alpha$, $\tilde r_F$, and $P_j$ invariant. Note that the degradation of magnetic order at the interface may significantly affect $P_\Sigma$ and $\tilde r_F$.

\section{Spin injection efficiency for a half-metallic electrode}\label{properties}

The conductance and reflectance parameters defined in Eq.\ (\ref{diffusion})-(\ref{diffusion2}) depend only on the properties of the interface region and not on the properties of the bulk half-metallic region attached to it. Therefore, the spin injection efficiency does not depend on the thickness of the half-metallic electrode. This result is valid as long as the half-metallic region is not so thin as to violate the assumptions of the diffusion theory. In practice this means that it should be thick compared to the electronic mean-free path.

Using this property, we can formally include an arbitrarily thick half-metallic layer in the definition of the interfacial region, so that the conductances $\Sigma_{\uparrow\uparrow}$, $\Sigma_{\uparrow\downarrow}$ are made very small (and thus the resistances $r_\uparrow$ and $r_\downarrow$ in Fig.\ \ref{resistors} very large). Then the first term in the right-hand side of (\ref{rfa}) is negligible compared to the corresponding asymptotic value $\tilde R^\infty_{\uparrow\downarrow}$, which is always finite at finite temperature. Thus, efficient spin-injection is possible only when $\tilde R^\infty_{\uparrow\downarrow}\lesssim r^{-1}_N$. In other words, spin injection is suppressed at $r_N\gg (\tilde R^{\infty}_{\uparrow\downarrow})^{-1}=r^\infty_{\uparrow\downarrow}$. On one hand, this condition is similar to the conductivity mismatch, because it sets a certain limit for $r_N$. On the other hand, the physical picture is quite different, because this limit is not related to the conductivity of the half-metal.

For generality let us allow for the existence of a magnetically unpolarized resistive barrier (such as a tunnel junction) at the interface, and let $T_c$ denote the transmission probability across this barrier. From the first term in (\ref{rfa}) we can deduce $\tilde r_F\sim (\alpha G_N T_c)^{-1}$. Alternatively, we can find $\tilde R^{\infty}_{\uparrow\downarrow}$ by adding a large resistor on the left of $T_c$ with transmission probability $T_F\ll T_c$. In order to reflect with a spin-flip, an electron incident from the N side must first tunnel across the barrier in order to reach the spin-disordered region; this gives a factor $T_c$. Since we require $T_F/T_c\to0$, the electron is reinjected back into the N region with probability 1. The probability of spin flip adds a factor $\alpha$, so we obtain $\tilde R^{\infty}_{\uparrow\downarrow}\sim \alpha G_N T_c$ and confirm the above result for $\tilde r_F$. An electron can also scatter with a spin flip on the transverse exchange field introduced by the spin density penetrating across the barrier. This mechanism contributes in the same order to $\tilde R^{\infty}_{\uparrow\downarrow}$.

In the Ohmic regime (low-resistance interface with $T_c\sim1$), we have $G_N\tilde r_F\sim1/\alpha$. Since $G_Nr_N\sim l^N_{sf}/l$, where $l$ is the mean-free path in the normal metal, we find that spin disorder suppresses spin injection when $\alpha\gg l/l^N_{sf}$.  At small $T$ we expect $\alpha\approx\langle\theta^2\rangle/4$, where $\theta$ is the polar angle of the injected spinor. The parameter $\alpha$ is approximately related to the reduced magnetization $m=M(T)/M(0)$ of the half-metal as $2\alpha\approx1-m$. (This quantity is proportional to the partial minority-spin DOS in the global spin basis.) Thus, the above condition shows the range of temperatures for which Ohmic spin injection from a half-metal may be possible.

It is interesting to compare this result with the case of a two-channel ferromagnet with the same spin polarization of DOS, for which the efficiency of Ohmic spin injection is given by Eq.\ (\ref{Pj}) with $r_c=0$. Setting $\rho_\downarrow\sim\rho_\uparrow/\alpha$, we find $r_F\sim \rho_\uparrow l_{sf}^F/\alpha$, which should be compared to $\tilde r_F\sim\rho_N l/\alpha$ in the case of a half-metallic electrode. The dependence on the spin polarization of the globally defined DOS is similar, but the overall factor is different: the product $\rho_\uparrow l_{sf}^F$ is replaced by $\rho_N l$ in the case of a half-metal. For spin injection in semiconductors from metals, typically $\rho_N\gg \rho_\uparrow$, while $l_{sf}^F$ is usually fairly small.\cite{Bass} Thus, the effective resistance of a half-metallic electrode may be much larger compared to a conventional ferromagnet with a similar spin polarization of the DOS, which is an advantage for practical applications. On the other hand, since $\tilde r_F$ does not depend on the resistivity of the half-metal, there is no benefit in increasing this resistivity. In particular, a magnetic semiconductor should not necessarily be a better spin injector than a highly conductive half-metal, assuming that half-metallicity is maintained at the interface in both cases.

For an actual device (e.\ g.\ F/N/F) it is necessary that $l^N_{sf}$ is not small compared to the length $L$ of the channel, and $l/l^N_{sf}$ should be replaced by $l/L$ if $l^N_{sf}\gg L$. Thus, for Ohmic spin injection from a half-metal it may be beneficial to use a lightly doped semiconducting channel in order to maximize the mean-free path there. This is contrary to the conventional conductance mismatch considerations, according to which it is desirable to decrease $\rho_N$ by increasing the doping concentration. The mean-free path in lightly doped silicon may be as high as 30~nm at room temperature \cite{Weber}. For a short channel with $L\sim 300$~nm this would allow Ohmic spin injection at $\alpha\lesssim 0.1$. Although this is an order-of-magnitude estimate, at face value it allows spin injection for $M(T)/M(0)\gtrsim0.8$. In elemental transition metals the reduced magnetization drops to 0.8 at about 75\% of the Curie temperature, so this limitation is not very restrictive, particularly since the half-metallic gap may be closed by magnetic disorder at much lower temperatures.\cite{Lezaic} A similar estimate applies to nonlocal spin valves with a copper channel, where at room temperature the mean-free path is on the order of 30~nm, and the spin-diffusion length is a few hundred nanometers.\cite{Bass} It is possible that efficient spin injection across a transparent interface observed in Ref.\ \onlinecite{Ramsteiner} can be understood in a similar way.

From the point of view of interface engineering, it is always necessary to avoid the depletion region near the surface.\cite{Jansen}
On the other hand, we would like to point out that in the Ohmic regime the existence of interface states in the ``wrong'' spin channel does not necessarily preclude spin injection, as it would with a tunnel barrier. If these states are strongly hybridized with the normal region, they can be regarded as a part of the corresponding spin channel. (See Ref.\ \onlinecite{Mavropoulos} for a related discussion.) It is, however, important that the magnetic continuity and ordering at the interface is maintained. Otherwise, partially ordered regions or ``loose spins'' can provide strong spin-flip scattering at relatively low temperatures, thereby violating the $\alpha\lesssim l/l^N_{sf}$ inequality and suppressing spin injection. Thus, interface design based on chemical similarity of the F and N regions \cite{Chadov} may in practice be counterproductive, because it may be expected to facilitate interdiffusion.

With a tunnel barrier at the interface, the temperature range allowing efficient spin injection extends to $\alpha\lesssim (l/l^N_{sf})/T_c$. As in the case of a conventional ferromagnetic electrode, the tunnel barrier is favorable for spin injection. In the case $T_c\ll l/l^N_{sf}$ spin injection is possible at any temperature, and its efficiency is proportional to $P_t=(1-\alpha)/(1+\alpha)$. (Of course, this assumes that the half-metal continues to behave as a single-channel conductor at elevated temperatures; real materials with a small half-metallic gap do not necessarily behave in this way.)

\section{Tight-binding calculations}
\label{TB}

\begin{figure}
\begin{center}
\includegraphics[width=0.35\textwidth]{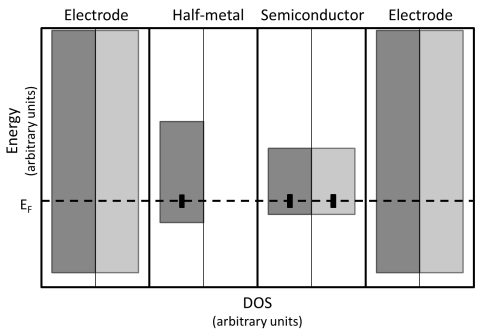}
\end{center}
\caption{Schematic picture of the band alignment for the spin injection device without spin disorder. Darker and lighter bands correspond to majority and minority spins. The minority-spin band in the half-metallic region is shifted up beyond the energy range shown in the figure. The horizontal dashed line shows the Fermi level, and the vertical black bars show the amplitude of random disorder.} \label{pic-fig1}
\end{figure}

In this section we verify the conclusions of the general theory using tight-binding calculations for a specific realization of a half-metal based on the $s$-$d$ model. Static spin disorder is introduced by randomizing the directions of the exchange fields on different sites according to the mean-field distribution function corresponding to the specified value of the magnetization. The spin injection device is treated as an elastic system. We use a single-band Hamiltonian with nearest-neighbor interactions in the simple cubic lattice. For the half-metallic region, the energies in one spin channel are made very large in order to lift it far above the Fermi level. This is the limit of a large $s$-$d$ exchange integral. The half-metallic and normal regions are sandwiched between two non-magnetic leads. The hopping and band center parameters are selected as showin schematically in Fig.\ \ref{pic-fig1}. Calculations were performed for supercells with the $10\times10$ cross-section. The $2\times2$ conductance matrix $G_{ss'}$ was obtained using the Landauer-B\"{u}ttiker approach, averaging over 100 configurations of spin and Anderson disorder. Brillouin zone integration was performed using a $5 \times 5$ mesh, which was sufficient for convergence. To simulate diffusive transport, random disorder was applied to both half-metallic and semiconductor regions. Fig.~\ref{pic-fig1} illustrates the relative amplitude of random disorder relative to the band widths.

Since the spin current in the normal region is conserved, the spin polarization of the current flowing into the right electrode is identified with the spin injection efficiency:
\begin{align}
  \label{eqn-spinpolarization}
P_G = \frac{G_{\uparrow\uparrow}+G_{\downarrow\uparrow} - G_{\uparrow\downarrow}-G_{\downarrow\downarrow}}{G_{\uparrow\uparrow}+G_{\downarrow\uparrow} + G_{\uparrow\downarrow}+G_{\downarrow\downarrow}}.
\end{align}
According to the results of the previous sections, the dependence of $P_G$ on the total resistance of the normal layer reflects the dependence of spin injection efficiency on the effective resistance $r_N$.

In the following we verify the following properties of the spin injection efficiency $P_G$ for a half-metallic electrode: (1) Independence of $P_G$ on the thickness of the half-metal; (2) The form (\ref{pjfromt}) of the dependence of $P_G$ on the resistance of the normal region; (3) Dependence of $\tilde r_F$ on the magnetization of the half-metal, $\tilde r_F\sim\alpha^{-1}$; (4) Dependence of $\tilde r_F$ on the transparency of a thin tunnel barrier inserted at the interface, $\tilde r_F\sim T_c^{-1}$.

\begin{figure}
\begin{center}
\includegraphics[width=0.4\textwidth]{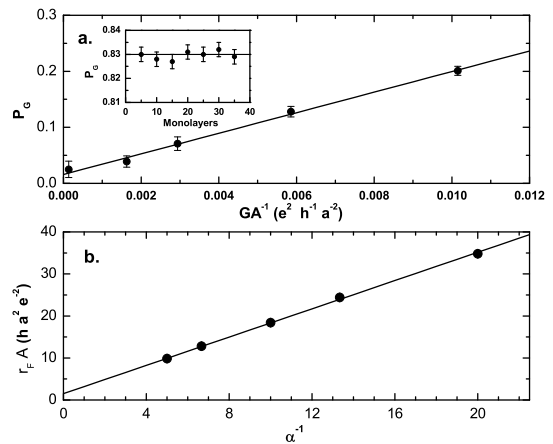}
\end{center}
\caption{\textbf{a.} Conductance polarization as a function of the inverse resistance-area product for the reduced magnetization $m=0.8$. Inset: Same quantity as a function of the thickness of the half-metal at $m=0.9$. \textbf{b.} Effective resistance $\tilde{r}_{F}$ as a function of $\alpha^{-1}$, where $\alpha=(1-m)/2$.} \label{pic-fig2}
\end{figure}

The inset of Fig.~\ref{pic-fig2}a shows $P_G$ as a function of the thickness of the half-metallic region at $m = M/M(0)=0.9$, with a 50-monolayer thick normal region. It is seen that $P_G$ is independent of the thickness of the half-metal, in agreement with the general results. In all subsequent calculations the thickness of the half-metal is fixed at 5 monolayers.

In order to obtain the asymptotic dependence of $P_G$ on the resistance of the normal region, we have added a tunnel barrier of variable height and thickness between the semiconductor and the right lead. This is necessary, because otherwise the localization effects become important when the disordered normal region is made too long,\cite{Todorov} and the diffusive scaling breaks down.
The length of the normal region is fixed at 50 monolayers. Fig.~\ref{pic-fig2}a shows the dependence of $P_G$ on the total resistance of the scattering region (which is dominated by the auxiliary tunnel barrier) for the reduced magnetization of $m = M/M(0) =0.8$. It is clearly seen that $P_G$ goes to zero linearly with $r_N^{-1}$, in agreement with Eq.\ (\ref{pjfromt}).

Next we evaluate the dependence of the effective resistance $\tilde{r}_F$ on the magnetization $m$ of the half-metal. To this end, for a given $m$ the $P_G$ is calculated from the configurationally averaged spin-dependence conductances for a set of thicknesses of the normal region ranging from 10 to 200 monolayers (all within the range where weak localization effects are undetectable). The $P_G(r_N)$ dependence is then fitted to Eq.\ (\ref{pjfromt}) for each $m$. The magnitude of $P_t$ decreases with decreasing magnetization but always remains somewhat larger than $m$. (This is likely due to the fact that the conduction electrons sample spin disorder over a few sites, effectively decreasing the transverse fields.) The dependence of $\tilde{r}_F$ on the magnetization is shown in Fig.~\ref{pic-fig2}b, where $\alpha = (1-m)/2$, as above. Linear dependence $\tilde{r}_F \propto \alpha^{-1}$ confirms the predictions of the general model, in which $\tilde r_F\propto R^{-1}_{\uparrow\downarrow}\sim\alpha^{-1}$.
This divergence of $\tilde r_F$ at low temperatures may be used experimentally as a signature of a true single-channel half-metal, although at very low temperatures the interfacial spin-orbit scattering may take over and cut off the divergence.

\begin{figure}
\begin{center}
\includegraphics[width=0.4\textwidth]{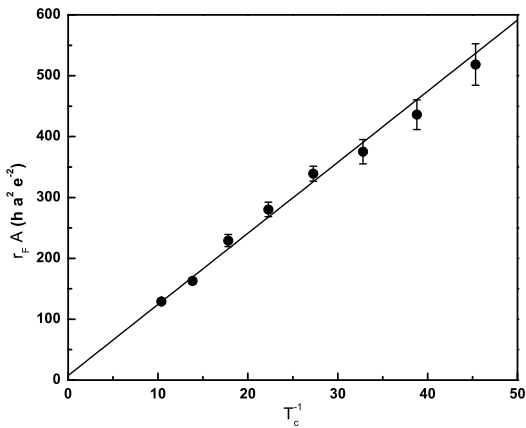}
\end{center}
\caption{Effective resistance $\tilde{r}_{F}$ as a function of $T_c^{-1}$ for fixed magnetization $m=0.6$.} \label{pic-fig3}
\end{figure}

Finally, we considered the effect of a single-monolayer tunnel barrier at the F/N interface. We set $m=0.6$ for the half-metal and varied the thickness of the semiconductor region from 10 to 200 monolayers, as above. As above, $P_G$ calculated from the averaged spin-dependent conductances was fitted to Eq.\ (\ref{pjfromt}), extracting the $\tilde r_F(T_c)$ dependence. $T_c$ was varied by changing the height of the tunnel barrier. Fig.\ \ref{pic-fig3} shows the results suporting the inverse relationship $\tilde r_F\propto T_c^{-1}$. Together with the results shown in Fig.\ \ref{pic-fig2}, we find $\tilde r_F\propto(\alpha T_c)^{-1}$, as expected.

\section{Conclusions}\label{concl}

We have analyzed the spin injection from a half-metallic electrode into a normal (or semiconducting) region in the presence of thermal spin disorder. The two-current model with independent populations of the two spin channels is inapplicable to a half-metallic ferromagnet. The spin injection efficiency is described by Eq.\ (\ref{pjfromt}), in which $r_N$ is the conventional effective resistance of the normal metal, while $\tilde r_F$ is controlled by spin-flip scattering at the interface with the ferromagnet. Although $\tilde r_F$ does not depend on the thickness of the half-metallic layer, its dependence on the spin polarization of the density of states and on the contact resistance is similar to the case of a conventional ferromagnet. Explicit simulations for the tight-binding $s$-$d$ model confirm these general conclusions. In the case of a transparent interface, efficient spin injection is possible in the temperature range corresponding to $\alpha\lesssim l/l^N_{sf}$, where $\alpha$ is the relative deviation of the (surface) magnetization from saturation, and $l$ is the mean-free path in the N region. A rough estimates suggests that efficient spin injection from half-metallic Co-based Heusler alloys into silicon or copper may be possible at room temperature across a transparent interface. Adding a tunnel barrier at the interface with transparency $T_c$ extends this temperature range to $\alpha\lesssim l/l^N_{sf}T^{-1}_c$.

\section*{Acknowledgments}

We are grateful to Andre Petukhov, Evgeny Tsymbal, and Igor \v Zuti\'c for useful discussions. This work was supported by NSF DMR-1005642 and by Research Corporation through a Cottrell Scholar Award (K.\ B.).

\end{document}